\documentclass[aps,showpacs,preprintnumbers,amsmath,amssymb,nofootinbibt,preprint]{revtex4}
\usepackage{graphicx}
\usepackage{amsmath}
\usepackage{amssymb}
\usepackage{amsfonts}
\usepackage{bm}
\usepackage{color}

\def\be{\begin{equation}}
\def\ee{\end{equation}}
\def\bea{\begin{eqnarray}}
\def\eea{\end{eqnarray}}

\begin{document}

\title{Condensate dark matter stars}
\author{X. Y. Li}
\email{lixinyu@hku.hk}
\affiliation{Department of Physics and
Center for Theoretical and Computational Physics, The University
of Hong Kong, Pok Fu Lam Road, Hong Kong, P. R. China}
\author{T. Harko}
\email{harko@hkucc.hku.hk}
\affiliation{Department of Physics and
Center for Theoretical and Computational Physics, The University
of Hong Kong, Pok Fu Lam Road, Hong Kong, P. R. China}
\author{K. S. Cheng}
\email{hrspksc@hkucc.hku.hk}
\affiliation{Department of Physics and
Center for Theoretical and Computational Physics, The University
of Hong Kong, Pok Fu Lam Road, Hong Kong, P. R. China}

\begin{abstract}
We investigate the structure and stability properties of compact astrophysical objects that may be formed from the Bose-Einstein condensation of dark matter. Once the critical temperature  of a  boson gas is less than the critical temperature,  a Bose-Einstein Condensation process can always take place during the cosmic history of the universe. Therefore we model the dark matter inside the star as a Bose-Einstein condensate, which can be described by a polytropic equation of state. We derive the basic general relativistic equations describing the equilibrium structure of the condensate dark matter star with spherically symmetric static geometry. The structure equations of the condensate dark matter stars are studied numerically. The critical mass and radius of the dark matter star are given by
$M_{crit}\approx 2(l_a/1fm)^{1/2}(m_{\chi}/1\;{\rm GeV})^{-3/2}M_{\odot}$ and $R_{crit}\approx 1.1 \times 10^6(l_a/1\;{\rm fm})^{1/2}(m_{\chi}/1\;{\rm GeV})^{-3/2}$ cm respectively, where
$l_a$ and $m_{\chi}$ are the scattering  length and the mass of dark matter particle, respectively.
\end{abstract}


\maketitle

\section{Introduction}

Since in the terrestrial experiments Bose-Einstein Condensation (BEC) is a well-known phenomenon, the possibility that a similar condensation may have occurred during the cosmological evolution of the universe cannot be excluded {\it a priori}. In fact, once the critical temperature  of the boson gas is less than the critical temperature,  BEC can always take place at some moment during the cosmic history of the universe.  Different aspects of the BEC cosmological transition were considered in \cite{Fuk05}-{\cite{Fuk09}. The
critical temperature for the condensation to take place is $T_{\rm cr}<2\pi \hbar
^{2}n^{2/3}/m_{\chi }k_{B}$, where $n$ is the particle number density, $m_{\chi }$ is the particle mass, and $k_B$ is Boltzmann's constant \cite{Da99}-\cite{Pet}. Since the matter temperature $T_m$ varies as $T_m\propto  a^{-2}$, where $a$ is the scale factor of the universe, it follows that during an adiabatic evolution the ratio of the photon temperature $T_{\gamma }$ and of the matter temperature evolves as $T_{\gamma }/T_m\propto a$.  Cosmic evolution has the same temperature dependence, since
in an adiabatic expansion process the density of a matter dominated universe evolves as $\rho \propto T^{3/2}$ \cite{Fuk08,Fuk09}. Therefore, if the boson temperature is equal, for example, to the radiation temperature at a redshift $z = 1000$,  the critical temperature for the Bose-Einstein Condensation is at present $T_{\rm cr} = 0.0027K$ \cite{Fuk09}.
On the other hand,  we expect that the universe is always under critical temperature, if it is at the present time \cite{Fuk09}.

It has been proposed recently that the first stars to
exist in the universe were powered by dark matter heating rather than by fusion \cite{ds1}-\cite{ds3}. Weakly Interacting
Massive Particles, collect inside the first stars and annihilate,
to produce a heat source that can power the stars. A new stellar phase results, called a Dark Star, powered
by dark matter annihilation as long as there is dark matter fuel. The heat source can power the Dark Stars for millions to billions of years. These objects can grow to be supermassive dark stars (SMDS) with masses $\sim (10^5-10^7) M_{\odot}$ \cite{ds3}. However it is not known what is the ratio between
dark matter and anti-dark matter or if the dark matter particle should even have an anti-particle partner to annihilate to generate the heat.
On the other hand non-annihilating dark matter particles can also generate very important effects in main sequence stars. In \cite{iocco} it was shown that the
energy transport mechanism induced by dark matter particles can produce unusual conditions in the core of main sequence stars, which constrain the
spin-dependent cross section $\geq 10^{-37}cm^2$, and the dark matter particle mass $m_{\chi}\geq 5$GeV.

It is the purpose of the present paper to consider a systematic study of the effects of the presence of condensate dark matter on the properties of compact astrophysical objects. By considering the condensation process in a cosmological setting, as a first step in our analysis we determine the physical and cosmological parameters at which the condensation of a boson gas can take place. If the critical temperature  of the gas is less than the critical temperature, the Bose-Einstein Condensation process takes place, resulting in the formation of a condensate dark matter background, which can be either accreted by normal matter baryonic stars, or can form stellar type objects due to gravitational instabilities.  Therefore we model the dark matter inside the star as a Bose-Einstein condensate.
The condensate dark matter equation of state  can be described by a polytropic equation of state, with polytropic index $n=1$. We derive the basic general relativistic equations describing the equilibrium structure of the condensate dark matter star with spherically symmetric static geometry, and we study the structure equations numerically. Our results show that the critical mass and radius of the condensate dark matters star are given by
$M_{crit}\approx 2(l_a/1fm)^{1/2}(m_{\chi}/1\;{\rm GeV})^{-3/2}M_{\odot}$ and $R_{crit}\approx 1.1 \times 10^6(l_a/1\;{\rm fm})^{1/2}(m_{\chi}/1\;{\rm GeV})^{-3/2}$ cm, respectively, where
$l_a$ and $m_{\chi}$ are the scattering  length and the mass of the dark matter particle.

The present paper is organized as follows. In Section \ref{sect2} the equations of state of the condensate and non-condensate dark matter are written down. The structure equations describing the properties of the general relativistic dark stars are derived in Section \ref{Sect3}. The results of the numerical integration of the structure equations are presented in Section \ref{Sect4}. We discuss and conclude our results in Section \ref{Sect5}.

\section{Dark  matter equations of state}\label{sect2}

\subsection{Bose-Einstein condensed dark matter}\label{2}

At very low temperatures, all particles in a dilute Bose gas condense to the
same quantum ground state, forming a Bose-Einstein Condensate (BEC). Particles become correlated with each other when their wavelengths
overlap, that is, the thermal wavelength $\lambda _{T}$ is greater than the
mean inter-particles distance $l$. This happens at a temperature $T_{\rm cr}\approx 2\pi \times
\hbar ^{2}\rho ^{2/3}/m^{5/3}k_{B}$, where $m$ is the mass of the particle in the
condensate, $\rho $ is the  density, and $k_{B}$ is Boltzmann's constant
\cite{Da99}. A coherent state develops when the particle density is enough
high, or the temperature is sufficiently low.
We assume that the dark matter halos are composed of a strongly - coupled dilute Bose-Einstein
condensate at absolute zero. Hence almost all the dark matter particles are
in the condensate.  In a dilute and cold gas,
only binary collisions at low energy are relevant, and these collisions are
characterized by a single parameter, the $s$-wave scattering length $l_a$,
independently of the details of the two-body potential. Therefore, one can
replace the interaction potential with an effective
interaction $V_I\left( \vec{r}^{\;\prime }-\vec{r}\right) =\lambda \delta \left(
\vec{r}^{\;\prime }-\vec{r}\right) $, where the coupling constant $\lambda $
is related to the scattering length $l_a$
through $\lambda =4\pi \hbar ^{2}l_a/m_{\chi }$ \cite{Da99}.
The ground state properties of the dark matter are
described by the mean-field Gross-Pitaevskii (GP) equation. The GP equation
for the dark matter halos can be derived from the GP energy functional,
\begin{eqnarray}
E\left[ \psi \right] &=&\int \left[ \frac{\hbar ^{2}}{2m_{\chi }}\left| \nabla \psi
\left( \vec{r}\right) \right| ^{2}+\frac{U_{0}}{2}\left| \psi \left( \vec{r}%
\right) \right| ^{4}\right] d\vec{r}-\frac{1}{2}Gm_{\chi }^{2}\int \int \frac{\left|
\psi \left( \vec{r}\right) \right| ^{2}\left| \psi \left( \vec{r}^{\;\prime
}\right) \right| ^{2}}{\left| \vec{r}-\vec{r}^{\;\prime }\right| }d\vec{r}d%
\vec{r}^{\;\prime }=\nonumber\\
&&E_K+E_{int}+E_{grav},
\end{eqnarray}
where $\psi \left( \vec{r}\right) $ is the wave function of the condensate,
and $U_{0}=4\pi \hbar ^{2}l_a/m_{\chi }$ \cite{Da99}. The first term in the energy functional is
the kinetic energy, the second is the interaction energy, and the third is
the gravitational potential energy. The mass density of the condensate dark matter is
defined as
\begin{equation}
\rho _{\chi }\left( \vec{r}\right) =m_{\chi }\left| \psi \left( \vec{r}\right)
\right| ^{2}=m_{\chi}\rho \left(\vec{r},t\right),
\end{equation}
and the normalization condition is $N=\int \left| \psi \left(
\vec{r}\right) \right| ^{2}d\vec{r}$, where $N$ is the total number of dark
matter particles. The variational
procedure  $\delta E\left[ \psi \right] -\mu \delta \int \left| \psi \left(
\vec{r}\right) \right| ^{2}d\vec{r}=0$ gives the GP equation as
\begin{eqnarray}\label{poi}
&&-\frac{\hbar ^{2}}{2m_{\chi }}\nabla ^{2}\psi \left( \vec{r}\right) +m_{\chi }V\left( \vec{r}%
\right) \psi \left( \vec{r}\right) + U_{0}\left| \psi \left( \vec{r}\right)
\right| ^{2}\psi \left( \vec{r}\right) = \mu \psi \left( \vec{r}\right) ,
\end{eqnarray}
where $\mu $ is the chemical potential, and the gravitational potential $V$
satisfies the Poisson equation
\be
\nabla ^{2}V=4\pi G\rho .
\ee
In the time-dependent case the generalized Gross-Pitaevskii equation describing a
gravitationally trapped rotating Bose-Einstein condensate is given by
\begin{eqnarray}\label{sch}
i\hbar \frac{\partial }{\partial t}\psi \left( \vec{r},t\right)=
\left[ -%
\frac{\hbar ^{2}}{2m_{\chi }}\nabla ^{2}+m_{\chi }V\left(
\vec{r}\right) +U_0\left| \psi \left( \vec{r},t\right)
\right| ^{2} \right] \psi \left( \vec{r},t\right) .
\end{eqnarray}

The physical properties of a Bose-Einstein condensate described by the
generalized Gross-Pitaevskii equation given by Eq.~(\ref{sch}) can be
understood much easily by using the so-called Madelung representation of the
wave function \citep{Da99,rev,Pet}, which consist in writing $\psi $ in the form
\begin{equation}
\psi \left( \vec{r},t\right) =\sqrt{\rho \left( \vec{r},t\right) }\exp \left[
\frac{i}{\hbar }S\left( \vec{r},t\right) \right] ,
\end{equation}
where the function $S\left( \vec{r},t\right) $ has the dimensions of an
action. By substituting the above expression of  $\psi \left( \vec{r}%
,t\right) $ into Eq.~(\ref{sch}), it decouples into a system of two
differential equations for the real functions $\rho _{\chi }$ and $\vec{v}$,
given by
\begin{equation} \label{cont}
\frac{\partial \rho _{\chi }}{\partial t}+\nabla \cdot \left( \rho _{\chi }\vec{v}%
\right) =0,
\end{equation}
\begin{eqnarray}
\rho _{\chi }\left[ \frac{\partial \vec{v}}{\partial t}+\left( \vec{v}\cdot
\nabla \right) \vec{v}\right] = -\nabla P_{\chi }\left(\frac{\rho _{\chi }}{m_{\chi }}\right)
 -\rho _{\chi }\nabla \left(
\frac{V}{m_{\chi }}\right) -\nabla V_{Q},  \label{euler}
\end{eqnarray}
where we have introduced the quantum potential
$V_{Q}=-\left(\hbar ^{2}/2m_{\chi }\right)\nabla ^{2}\sqrt{\rho _{\chi }}/\sqrt{\rho
_{\chi }}$, and the velocity of the quantum fluid $\vec{v}=\nabla S/m_{\chi }$,
respectively. The effective pressure of the condensate is given by
\begin{equation}\label{pres2}
P_{\chi }\left( \frac{\rho _{\chi }}{m_{\chi }}\right) =u_{0 }\rho _{\chi }^{2},
\end{equation}
where
\begin{equation}
u_{0}=\frac{2\pi \hbar ^2 l_a}{m_{\chi }^3}.
\end{equation}

The Bose-Einstein gravitational condensation can be described as a gas whose
density and pressure are related by a polytropic equation of state,  with index $n=1$ \cite{Da99,Pet}.
When the number of particles in the gravitationally bounded Bose-Einstein
condensate states becomes large enough, the quantum pressure term makes a
significant contribution only near the boundary of the condensation. Thus the quantum
stress term in the equation of motion of the condensate can be neglected.
This is the Thomas-Fermi approximation, which has been extensively used for
the study of the Bose-Einstein condensates \cite{Da99, rev, Pet}. As the number of
particles in the condensate becomes infinite, the Thomas-Fermi approximation
becomes exact. This approximation also corresponds to the
classical limit of the theory. From its definition it follows that the velocity field is irrotational,
satisfying the condition $\nabla \times \vec{v}=0$.

\subsection{Normal dark matter equation of state}

We assume that in the early stages of the evolution of the universe dark matter consisted of bosonic particles of mass $m_{\chi}$ and temperature $T$, originating in equilibrium and decoupling at a temperature $T_D$ or chemical potential $\mu >>m_{\chi }$. By assuming that the dark matter forms an isotropic gas of particles in kinetic equilibrium, the spatial number density is given by
\begin{equation}
n=\frac{g}{h^3}\int{4\pi f(p)p^2dp},
\end{equation}
where $h$ is Planck's constant, $g$ is the number of helicity states, and
\begin{equation}
f(p)=\left[\exp\left(E-\mu \right)-1\right]^{-1},
\end{equation}
where $p$ is the momentum of the particle and $E=\sqrt{p^2+m_{\chi }^2c^4}$ is the energy. A particle species that decouples in the early universe from the remaining plasma at temperature $T_D$ redshifts its momenta according to $p(t)=p_Da_D/a(t)$, where $a(t)$ is the cosmological scale factor and $a_D$ is the value of the scale factor at the decoupling. The number density of the particles $n$ evolves as $n_{\chi }\sim a^{-3}(t)$ \cite{mad}. The distribution function $f$ at a time $t$ after the decoupling is related to the value of the  distribution function at the decoupling by $f(p)=f\left(pa/a_D\right)$. The distribution function keeps an equilibrium shape in two regimes. In the extreme-relativistic case, when $E\approx pc$, $T=T_Da_D/a$, and $\mu =\mu _Da_D/a$, respectively, the distribution function is given by $f_{\rm ER}(p)=\left[\exp\left(pc-\mu \right)-1\right]^{-1}$. In the non-relativistic decoupling case $E-\mu \approx p^2/2m_{\chi }-\mu _{\rm kin}$, where we have defined $\mu _{\rm kin}\equiv \mu -m_{\chi }c^2$, the distribution function is $f_{\rm NR}(p)=\left[\exp\left(p^2/2m_{\chi }-\mu _{\rm kin} \right)-1\right]^{-1}$. In the non-relativistic case $\mu _{\rm kin}$ and $T$ evolve as $\mu _{\rm kin}=\mu _{\rm kin,D}\left(a_D/a\right)^2$ and $T=T_D/\left(a_D/a\right)^2$, respectively \cite{mad}.

The kinetic energy-momentum tensor $T^{\mu }_{\nu }$ associated to the frozen distribution of dark matter is given by
\begin{equation}
T^{\mu }_{\nu }=\frac{g}{3h^3}\int{\frac{p^{\mu }p_{\nu }}{p^0}f(p)d^3p}.
\end{equation}
The energy density $\epsilon $ of the system is defined as
\begin{equation}
\epsilon =\frac{g}{3h^3}\int{Ef(p)d^3p},
\end{equation}
while the pressure of a system with an isotropic distribution of momenta is given by
\begin{equation}
P=\frac{g}{3h^3}\int{pvf(p)d^3p}=\frac{g}{3h^3}\int{\frac{c^2p^2}{E}f(p)d^3p},
\end{equation}
where the velocity $v$ is related to the momentum by $v=pc^2/E$ \cite{shap}.
In the non-relativistic regime, when $E\approx m_{\chi }c^2$ and $p\approx m_{\chi}v_{\chi }$,   the density $\rho _{\chi }$ of the dark matter is given by $\rho _{\chi }=m_{\chi }n_{\chi }$, while its pressure $P_{\chi }$ can be obtained as \cite{mad}
\begin{equation}
P_{\chi}=\frac{g}{3h^3}\int{\frac{p^2c^2}{E}f(p)d^3p}\approx 4\pi \frac{g}{3h^3}\int{\frac{p^4}{m_{\chi }}dp},
\end{equation}
giving
\begin{equation}\label{pres1}
P_{\chi}=\rho _{\chi }c^2\sigma_v^2,
\end{equation}
where $\sigma_v^2=\langle \vec{v}^{\;2}_{\chi } \rangle /3c^2$, and $\langle \vec{v}^{\;2}_{\chi } \rangle$ is the average squared velocity of the particle. $\sigma_v $ is the one-dimensional velocity dispersion.

\subsection{The cosmological Bose-Einstein transition process}

In order to analyze the cosmological conditions for the formation of a dark matter Bose-Einstein condensate we follow the approach introduced in \cite{H1}. The cosmological parameters at which the Bose- Einstein condensation process took place can be estimated by taking into account that the laws of thermodynamics require
that both the chemical potential $\mu $ and the pressure $p$ are single valued functions, that is, for any given values of the particle number  $n$ and $T$, there must only exist a single value of $\mu $  or $p$, respectively \cite{ft}.

Therefore, a first thermodynamic condition that must be satisfied during the cosmological Bose-Einstein Condensation process is the continuity of the pressure at the transition point. With the use of Eqs.~(\ref{pres2}) and (\ref{pres1}) the continuity of the pressure uniquely fixes the critical transition density $\rho _{\chi }^{\rm cr}$ from the normal dark matter state to the Bose-Einstein condensed state as
\begin{equation}\label{crde}
\rho _{\chi }^{\rm cr}=\frac{c^2\sigma ^2}{u_0}=\frac{c^2\sigma ^2m_{\chi }^3}{2\pi \hbar ^2 l_a}.
\end{equation}

The numerical value of the transition density depends on three unknown parameters, the dark matter particle mass, the scattering length, and the dark matter particles velocity dispersion, respectively. By assuming a typical mass of the dark matter particle of the order of 1 GeV (1 GeV = $1.78\times 10^{-24}$ g), a typical scattering length of the order of 1 fm, and a mean velocity square of the order of $\langle\vec{v}^2\rangle=9\times 10^{14}\;{\rm cm^2/s^2}$, the critical transition density can be written as
\begin{eqnarray}
\rho _{\chi }^{\rm cr}&=&7.327\times 10^{9}\left(\frac{\sigma ^2}{10^{-6}}\right)\left(\frac{m_{\chi }}{1\;{\rm GeV}}\right)^3\left(\frac{l_a}{1\;{\rm fm}}\right)^{-1}\;{\rm g/cm^3}.
\end{eqnarray}

The critical temperature at the moment of Bose-Einstein condensate transition is given by \cite{Da99}-\cite{Pet}
\begin{eqnarray}
T_{\rm cr}&\approx &\frac{2\pi \hbar ^2}{\zeta (3/2)^{2/3}m_{\chi }^{5/3}k_B}\left(\rho _{\chi }^{\rm cr}\right)^{2/3}=\frac{\left(2\pi \hbar ^2\right) ^{1/3}c^{4/3}}{\zeta (3/2)^{2/3}k_B}\frac{\left(\sigma ^2\right)^{2/3}m_{\chi }^{1/3}}{l_a^{2/3}},
\end{eqnarray}
where $\zeta (3/2)$ is the Riemann zeta function, or
\begin{eqnarray}
T_{\rm cr}&\approx &3.76\times10^8\left(\frac{m_{\chi }}{1\;{\rm GeV}}\right)^{1/3}\left(\frac{\sigma ^2}{10^{-6}}\right)^{2/3}\left(\frac{l_a}{1\;{\rm fm}}\right)^{-2/3}\;\rm K.
\end{eqnarray}

The critical pressure of the dark matter fluid at the condensation moment can be obtained as
\begin{eqnarray}
P_{\rm cr}&=&6.51\times 10^{24}\left(\frac{\sigma ^2}{10^{-6}}\right)^2\left(\frac{m_{\chi }}{1\;{\rm GeV}}\right)^3\left(\frac{l_a}{1\;{\rm fm}}\right)^{-1}\;{\rm dyne/cm^2}.
\end{eqnarray}

By assuming that the Universe is flat, the cosmological dynamics of the dark matter before the condensation is described by the standard Friedmann-Robertson-Walker model, in which the evolution of the normal dark matter density is given by the equation
\begin{equation}
\dot{\rho _{\chi}}+3\rho _{\chi}(1+\sigma ^2)\frac{\dot{a}}{a} =0,
\end{equation}
where $a$ is the scale factor of the Universe, with the general solution given by
\begin{equation}
\rho _{\chi }=\frac{\rho _{\chi ,0}}{\left(a/a_0\right)^{3\left(1+\sigma ^2\right)}},
\end{equation}
where $\rho _{\chi ,0}$ is the density of the dark matter at $a=a_0$. Usually the scale factor is normalized so that at the present time $a=a_0=1$.
Therefore the  critical value $a_{cr}$ of the scale factor of the Universe at the moment of the beginning of the Bose-Einstein condensation can be immediately obtained as
\begin{eqnarray}
a_{cr}/a_0=\left(\frac{\rho _{\chi ,0}u_0}{c^2\sigma ^2}\right)^{1/3\left(1+\sigma ^2\right)}=
\left(\frac{2\pi \hbar ^2l_a \rho _{cr ,0}\Omega _{\chi ,0}}{c^2\sigma ^2m_{\chi }^3}\right)^{1/3\left(1+\sigma ^2\right)},
\end{eqnarray}
where we have introduced the critical density of the Universe $\rho _{cr ,0}=3H_0^2/8\pi G$, and the dark matter density
parameter $\Omega _{\chi ,0}=\rho _{\chi,0}/\rho _{cr,0}$, where $H_0$ is the present day Hubble constant.  Hence for the value of the cosmological redshift at which the Bose-Einstein transition did occur we obtain the expression
\begin{equation}
1+z_{cr}=\left(\frac{2\pi \hbar ^2l_a \rho _{cr ,0}\Omega _{\chi ,0}}{c^2\sigma ^2m_{\chi }^3}\right)^{-1/3\left(1+\sigma ^2\right)}.
\end{equation}

In the following
for the Hubble constant we adopt  the value $H_{0}=70\;{\rm km}/{\rm s}/{\rm Mpc}=2.27\times 10^{-18}\;{\rm s}^{-1}$, giving for the critical
density a value of $\rho _{cr,0}=9.24\times 10^{-30}\;{\rm g}/{\rm cm}^{3}$. The present day dark matter density
parameter is $\Omega _{\chi ,0}\approx 0.228$, respectively \cite{Hin09}. By using the adopted numerical values of the constants we obtain for the critical scale factor and for the critical redshift at which the Bose-Einstein condensation took place the values
 \begin{eqnarray}
\frac{a_{cr}}{a_0}&\approx &4.58\times10^{-14}\times\left(\frac{m_{\chi }}{1\;{\rm GeV}}\right)^{-\left(1+\sigma ^2\right)}
\left(\frac{\sigma ^2}{3\times 10^{-6}}\right)^{-1/3\left(1+\sigma ^2\right)}
\left(\frac{l_a}{1\;{\rm fm}}\right)^{1/3\left(1+\sigma ^2\right)},
\end{eqnarray}
and
\begin{eqnarray}
z_{cr}&\approx &2.171\times10^{11}\times\left(\frac{m_{\chi }}{1\;{\rm GeV}}\right)^{\left(1+\sigma ^2\right)}
\left(\frac{\sigma ^2}{3\times 10^{-6}}\right)^{1/3\left(1+\sigma ^2\right)}
\left(\frac{l_a}{1\;{\rm fm}}\right)^{-1/3\left(1+\sigma ^2\right)},
\end{eqnarray}
respectively.

These results show that if dark matter consists of self-interacting particles with masses of the order of 1 GeV, and scattering length of the order of 1 fm, the transition from normal dark matter to condensate dark matter did occur very early in the history of the Universe, presumably even during the post-inflationary reheating phase. These results are consistent with those obtained in \cite{Fuk09}, where the inflation
is naturally initiated by the condensation of the bosons in the huge vacuum energy.  Bose-Einstein condensation  can take place and continue provided the boson temperature is less than the critical temperature at some moment of
cosmic evolution. The corresponding critical boson mass for BEC can be arbitrary in general, and there are no theoretical restrictions for its value \cite{Fuk09}. After the beginning of the phase transition the density of the dark matter $\rho _{\chi }\left(
t\right) $ decreases from $\rho _{\chi }^{\rm cr}\left( T_{\rm cr}\right) \equiv
\rho ^{\rm nor}_{\chi}$ (when all the dark matter is in a normal, non-condensed form) to $\rho _{\chi }\left( T_{cr}\right) \equiv \rho ^{BEC}_{\chi}$, corresponding to the full conversion of dark matter into a condensed state.

\section{Hydrostatic equilibrium equations of  dark matter stars}\label{Sect3}

In the following we restrict our study of the condensate dark matter star to
the static and spherically symmetric case, with the metric  represented as
\begin{equation}\label{metr1}
ds^{2}=e^{\nu (r)}c^2dt^{2}-e^{\lambda (r)}dr^{2}-r^{2}\left( d\vartheta
^{2}+\sin ^{2}\vartheta d\phi ^{2}\right) .
\end{equation}

For the metric given by Eq.~(\ref{metr1}), the Einstein gravitational field
equations, describing the dark matter star, take the form~
\begin{eqnarray}
-e^{-\lambda }\left( \frac{1}{r^{2}}-\frac{\lambda ^{\prime }}{r}\right) +%
\frac{1}{r^{2}}=\frac{8\pi G}{c^4} \varepsilon ,  \label{f1} \\
e^{-\lambda }\left( \frac{\nu ^{\prime }}{r}+\frac{1}{r^{2}}\right) -\frac{1%
}{r^{2}}=\frac{8\pi G}{c^4} \sigma ,  \label{f2}\\
\frac{1}{2}e^{-\lambda }\left( \nu ^{\prime \prime }+\frac{\nu ^{\prime 2}}{2%
}+\frac{\nu ^{\prime }-\lambda ^{\prime }}{r}-\frac{\nu ^{\prime }\lambda
^{\prime }}{2}\right) =\frac{8\pi G}{c^4} \sigma ,  \label{f3}
\end{eqnarray}
and
\begin{equation}\label{f4}
\nu ^{\prime }=-2\frac{\sigma ^{\prime }}{\varepsilon +\sigma },
\end{equation}
where $^{\prime }=d/dr$, $\varepsilon $ is the total energy-density, and $\sigma$ is the pressure along the radial
direction. Equation (\ref{f4}) is the consequence of the
conservation of the energy-momentum tensor, $T^{\mu }{}_{\nu;\mu }=0$. Eq.~(\ref{f1}) can be easily integrated to give
\begin{equation}
e^{-\lambda }=1-\frac{2GM(r)}{c^2r},
\end{equation}
where $M(r)=4\pi \int \left(\varepsilon /c^2\right)r^{2}dr$. W\texttt{}ith the use of Eqs.~(\ref{f2}) and (\ref{f4}) we obtain the mass
continuity equation, and the generalized Tolman-Oppenheimer-Volkoff (TOV)
equation describing the spherical symmetric static dark matter
configurations,
\begin{equation}
\frac{dM}{dr}=4\pi \frac{\varepsilon }{c^2}r^{2},  \label{eqcont0}
\end{equation}
\begin{equation}
\frac{d\sigma }{dr}=-\frac{\left( \varepsilon +\sigma \right) \left[ \left(4\pi G/c^4\right)
\sigma r^{3}+GM/c^2\right] }{r^{2}\left( 1-2GM/c^2r\right) }.  \label{tov0}
\end{equation}

In the case of the dark matter stars $\varepsilon=\rho_{\chi}c^2$ and $\sigma=P_{\chi}=\left(2\pi \hbar ^2l_a/m_{\chi }^3\right)\rho _{\chi }^2$. In order to simplify the equations we introduce the dimensionless dark matter density $\theta _{\chi}$, defined as $\rho _{\chi}=\rho _{c\chi}\theta _{\chi}$, where $\rho _{c\chi}$ is the central density of the dark matter.

In addition, let us denote
\begin{equation}
\alpha _{\chi }=\frac{u_{0}\rho _{c\chi }}{c^{2}}=\frac{2\pi \hbar ^2 l_a}{c^2m_{\chi }^3}\rho _{c\chi }=1.658\left(\frac{l_a}{{\rm 1\;fm}}\right)\left(\frac{m_{\chi }}{m_p}\right)^{-3}\left(\frac{\rho _{c\chi}}{10^{16}\;{\rm g/cm^3}}\right),
\end{equation}
The coefficient $\alpha_{\chi}$ can be broken into two parts. Let $\alpha_{\chi}=\beta\rho_{c\chi}$, where
\begin{equation}\label{beta}
\beta=\frac{2\pi \hbar ^2 l_a}{c^2m_{\chi }^3}=1.658\times 10^{-16}\left(\frac{l_a}{{\rm 1\;fm}}\right)\left(\frac{m_{\chi }}{m_p}\right)^{-3}\rm cm^{3}/g.
\end{equation}
$\beta$ is only determined by the properties of dark matter particle. Furthermore, $\alpha_{\chi}$ can be constrained by the condition that the sound speed of condensate dark matter must be smaller than the speed of light. Since
\begin{equation}
P_{\chi}=\frac{\alpha_{\chi}}{\rho_{c\chi}}c^{2}\rho_{\chi}^{2},
\end{equation}
the constraint on sound speed $\sqrt{\frac{\partial P_{\chi}}{\partial \rho_{\chi}}}<c$ gives
$\alpha_{\chi}<\frac{\rho_{c\chi}}{2\rho_{\chi}}$. This condition must be valid everywhere inside the star, at the center of the center it gives
\begin{equation}
\alpha_{\chi}<\frac{1}{2}.
\end{equation}
Later we will find that $\alpha_{\chi max}=0.43$ from the instability analysis.

By introducing a set of dimensionless variables $\left( \eta ,m\right) $,
defined as
\begin{equation}\label{parr}
r=\frac{c}{\sqrt{4\pi G\rho _{c\chi }}}\eta ,
\end{equation}
and
\begin{equation}\label{parm}
M=\frac{1}{\sqrt{4\pi }}\left( \frac{c^{2}}{G}\right) ^{3/2}\frac{1}{\sqrt{%
\rho _{c\chi }}}m,
\end{equation}
the structure equations of the dark matter star can be written as
\begin{equation}
\frac{dm}{d\eta }=\theta_{\chi}\eta ^{2},  \label{seqcont1}
\end{equation}
\begin{equation}
2\alpha_{\chi}\theta_{\chi}\frac{d\theta_{\chi}}{d\eta }=-\frac{\left( \theta_{\chi}+\alpha_{\chi}\theta_{\chi}^{2}\right) \left[ \alpha_{\chi}\theta_{\chi}^{2}\eta ^{3}+m\right] }{\eta ^{2}\left(
1-2m/\eta \right) }.
\label{stov1}
\end{equation}

The initial conditions for the structure equations are
\begin{equation}
m(0)=0,
\end{equation}
\begin{equation}
\theta_{\chi}(0)=1.
\end{equation}
And boundary of the object is reached when $\theta_{\chi}=0$ at which point the values for $m$ and $\eta$ are denoted by $m_{0}$ and $\eta_{0}$.

\section{Numerical Results}\label{Sect4}

The solutions of the dimensionless equations (\ref{seqcont1}) and (\ref{stov1}) are studied numerically for some fiducial values $m_{\chi}=m_{p}$, $l_{a}=1\;\rm fm$, and $\rho_{c\chi}=10^{13}\rm g/cm^{3}$. The variation of the dimensionless density $\theta_{\chi}$ as a function of the dimensionless radius $\eta$ is plotted in Fig.~\ref{dms}.

\begin{figure}[htbp]
   \centering
   \includegraphics[width=4in]{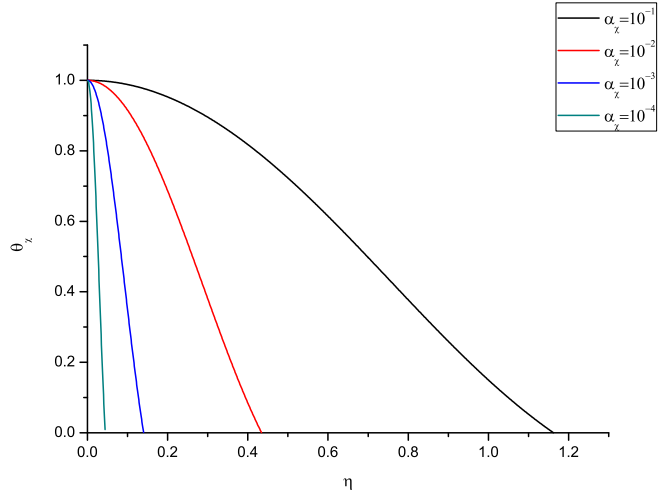}
   \caption{Dimensionless density $\theta_{\chi}$ versus dimensionless radius $\eta$ for condensate dark matter stars with different values of $\alpha_{\chi}$.
   }
   \label{dms}
\end{figure}
The only parameter in the structure equations of the star Eqs.~(\ref{seqcont1}) and (\ref{stov1}) is $\alpha_{\chi}$. The dimensionless mass $m_{0}$ and the dimensionless radius $r_{0}$ are thus functions of $\alpha_{\chi}$ only. The dependence of $m_{0}$ and $\eta_{0}$ on $\alpha_{\chi}$ can be fitted by simple relations valid for physical choice of $\alpha_{\chi}<1/2$
\begin{equation}\label{fm}
m_{0}=8.89\alpha_{\chi}^{3/2}\left(1+6\alpha_{\chi}\right)^{-4/3},
\end{equation}
\begin{equation}\label{fr}
\eta_{0}=4.45\alpha_{\chi}^{1/2}\left(1+6\alpha_{\chi}\right)^{-2/5}.
\end{equation}
Figs.~\ref{fitm} and \ref{fite} illustrate the comparison between the numerically obtained values and the fitted values as functions of $\alpha_{\chi}$. Fig.~\ref{relerr} illustrates the relative error of the fitted values from Eqs.~(\ref{fm}) and (\ref{fr}), and it shows that they are very good approximation formulae.
It can be seen that Eq.~(\ref{fm}) and Fig.~\ref{fr} give very good fitting to the numerically obtained values.

\begin{figure}[htbp] 
   \centering
   \includegraphics[width=4in]{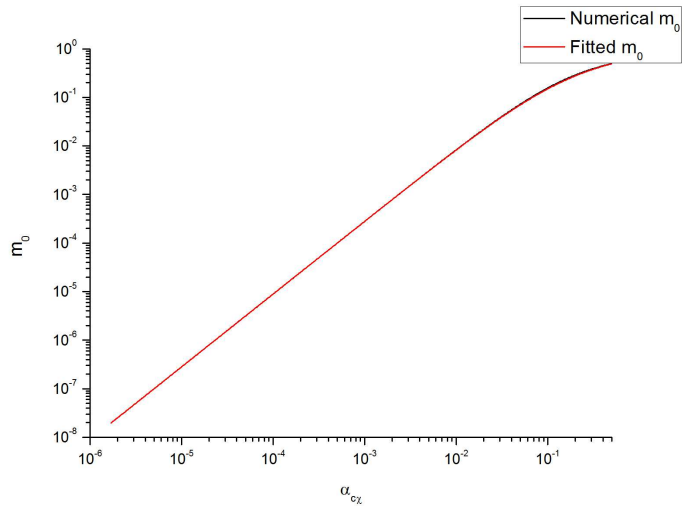}
   \caption{Comparison between the numerically obtained $m_{0}$ and the fitted $m_{0}$ from Eq.~(\ref{fm}): the black line is for the numerically obtained $m_{0}$, and the red line gives the fitted $m_{0}$ from Eq.~(\ref{fm}).}
   \label{fitm}
\end{figure}
\begin{figure}[htbp] 
   \centering
   \includegraphics[width=4in]{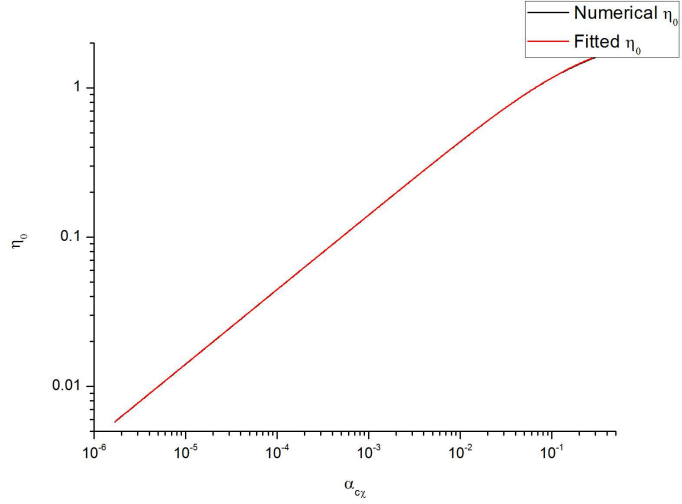}
   \caption{Comparison between the numerically obtained $\eta_{0}$ and the fitted $\eta_{0}$ from Eq.~(\ref{fr}): the black line represents the numerically obtained $\eta_{0}$, and the red line represents the fitted $\eta_{0}$ from Eq.~(\ref{fr}).}
   \label{fite}
\end{figure}
\begin{figure}[htbp] 
   \centering
   \includegraphics[width=4in]{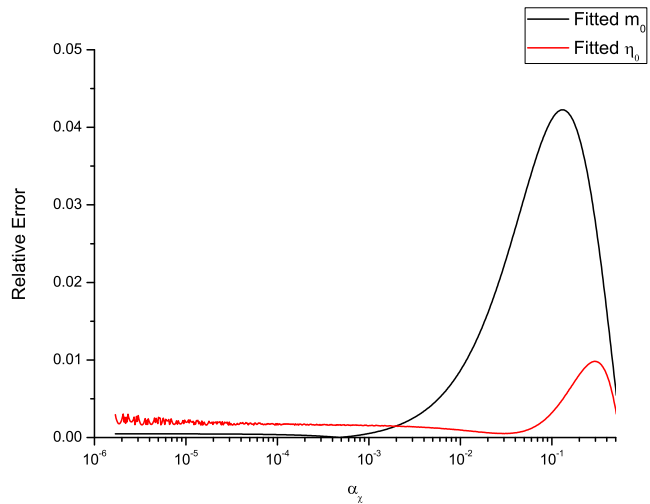}
   \caption{Relative error of the fitted $m_{0}$ and $\eta_{0}$ from Eqs.~(\ref{fm}) and (\ref{fr}): the black line represents $m_{0}$ and the red line represents $\eta_{0}$.}
   \label{relerr}
\end{figure}

The dimensionless mass $m_{0}$ increases with the dimensionless radius $\eta_{0}$, and there is no instability. However, when Eq.~(\ref{parm}) and Eq.~(\ref{parr}) are
used to calculate the physical mass and radius, they will behave quite differently. When the central density keeps increasing, there is a maximum mass corresponding to the instability. The instability is clearly seen in Fig.~\ref{pmr} for different choices of $\beta$ when $dM/dR=0$.
We will explain the origin of the instability in the next Section. Fig.~\ref{pma} illustrates the dependence of the physical mass on the parameter $\alpha_{\chi}=\beta\rho_{c\chi}$ for different values of  $\beta$.

There is a simple criteria for instability onset
\begin{equation}\label{crit}
\alpha_{\chi}=\beta\rho_{c\chi}=0.43,
\end{equation}
which is smaller than the constraint value obtained by the sound speed limit. In the next Section we will explain the origin of this value.

 By using Eq.~(\ref{fm}) and Eq.~(\ref{fr}), we can derive simple relations between the critical mass and radius corresponding to the instability, and the dark matter particles property parameter $\beta$,
\begin{equation}
M_{\rm crit}=1.54 M_{\odot}\left(\frac{\beta}{10^{-16}\rm cm^{3}/g}\right)^{\frac{1}{2}}\approx 2(l_a/1\;{\rm fm})^{1/2}(m_{\chi}/1\;{\rm GeV})^{-3/2}M_{\odot},
\end{equation}
and
\begin{equation}
R_{\rm crit}=8.75\times10^{5}\rm cm \left(\frac{\beta}{10^{-16}\rm cm^{3}/g}\right)^{\frac{1}{2}}\approx 1.1 \times 10^6(l_a/1\;{\rm fm})^{1/2}(m_{\chi}/1\;{\rm GeV})^{-3/2}\rm cm,
\end{equation}
respectively.

\begin{figure}[htbp] 
   \centering
   \includegraphics[width=4in]{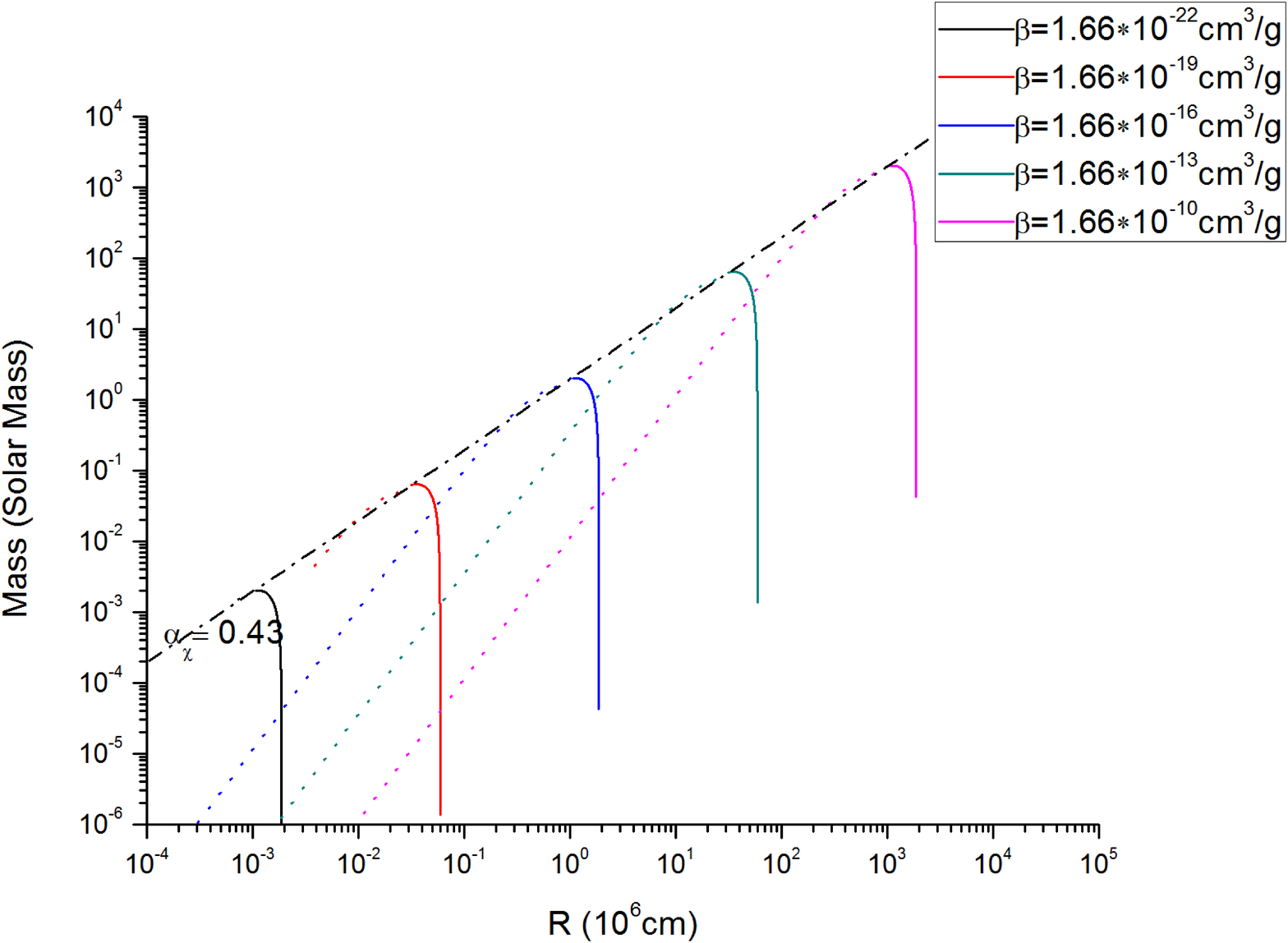}
   \caption{Numerical relation between physical mass and radius for different choices of $\beta$: the dashed lines correspond to non-physical condition where $\alpha_{\chi}>1/2$.}
   \label{pmr}
\end{figure}
\begin{figure}[htbp]
   \centering
   \includegraphics[width=4in]{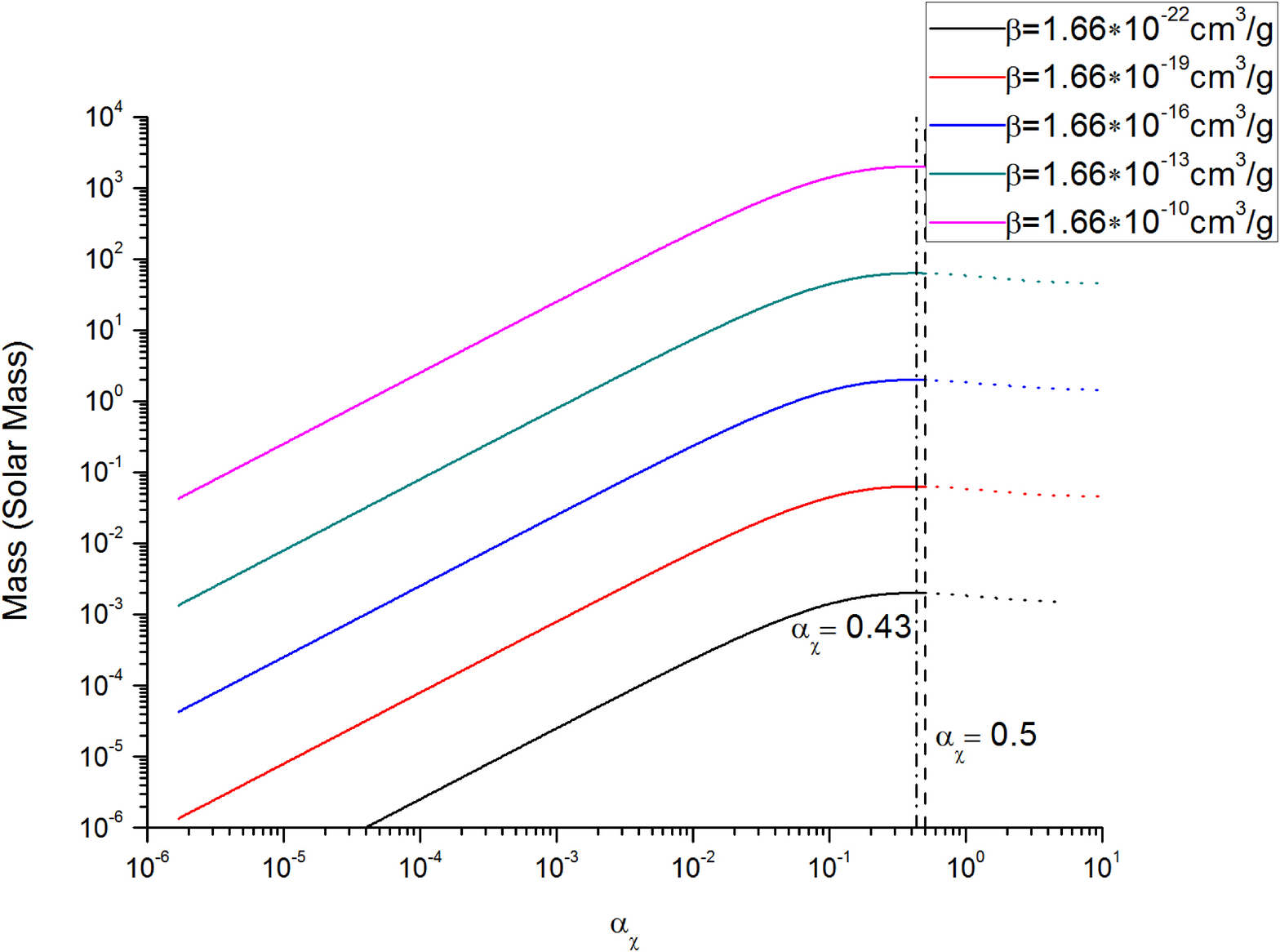}
   \caption{Numerical relation between physical mass and $\alpha_{\chi}$ for different choices of $\beta$: the dashed lines correspond to non-physical condition where $\alpha_{\chi}>1/2$.}
   \label{pma}
\end{figure}

\section{Discussions and final remarks}\label{Sect5}

In the present study, we have shown that it is possible to form stable dark matter stellar objects, given that dark matter has self-interaction determining the condensation process. The mass and the radius of the star depend on two parameters, one being an "environmental" parameter, the central density $\rho_{c\chi}$, and the other is the intrinsic property of dark matter particle $\beta$.

The parameter $\beta$ is a combination of dark matter particle mass $m_{\chi}$ and the self-interaction scattering length $l_{a}$, which is related to the self-interaction cross section $\sigma_{\chi\chi}$ by the relation $\sigma_{\chi\chi}=4\pi l_{a}^{2}$. Currently, the most popular dark matter model, the Weakly-Interacted-Massive-Particle (WIMPs) model favors a mass range between GeV to several TeV.  A possible dark matter candidate particles, the sterile neutrinos, have a mass in the keV range. This mass range cannot explain the cold dark matter structure formation. Other models of dark matter, like supersymmetric particles, light gravitons, axions predict different mass ranges (cf. \cite{feng}, \cite{bert}). The idea that dark matter may have self interaction was first proposed by Spergel and Steinhardt \cite{st00} to alleviate several apparent conflicts between astrophysical observations and the collisionless CDM model. Their proposed self-interaction has the strength range $\sigma_{\chi\chi}/m_{\chi}=0.45-450\;\rm cm^{2}/g$. The experimental constraint of self-interaction cross section mostly come from astronomical observation of colliding Bullet Cluster with the upper limit $\sigma_{\chi\chi}/m_{\chi}<1\;\rm cm^{2}/g$ \cite{mark03}. The condensate dark matter star might be able to provide another means to constrain the properties of dark matter particles.

The simple criteria for instability, Eq.~(\ref{crit}), might have a simple explanation by calculating the total energy of the condensate dark matter star. The total energy is the sum of negative gravitational energy $E_{\rm grav}$ and the internal energy $E_{\rm int}$. By Newtonian approximation the gravitational energy is
\begin{equation}
E_{\rm grav}=\int-\frac{GM(r)}{r}\mathrm{d}m=-\frac{c^{5}}{G\sqrt{4\pi G \rho_{c\chi}}}\int m\theta_{\chi}\eta\mathrm{d}\eta,
\end{equation}
and the internal energy of the condensate dark matter is
\begin{equation}
E_{\rm int}=\int u_{0}\rho_{\chi}^{2}\mathrm{d}V=\alpha_{\chi}\frac{c^{5}}{G\sqrt{4\pi G\rho_{c\chi}}}\int\theta_{\chi}^{2}\eta^{2}\mathrm{d}\eta.
\end{equation}
At the instability point, $E=E_{\rm grav}+E_{\rm int}=0$. This simplifies to be
\begin{equation}
\alpha_{\chi}\int\theta_{\chi}^{2}\eta^{2}\mathrm{d}\eta-\int m\theta_{\chi}\eta\mathrm{d}\eta=0.
\end{equation}
Since the two integrals are only functions of $\alpha_{\chi}$, the instability onset corresponds to the zero point of an equation of $\alpha_{\chi}$, which leads to Eq.~(\ref{crit}).

When a condensate dark matter is formed, it will lead to the accretion of material from space. The incoming dark matter particles will interact with the dark matter in the condensate star, transfer their kinetic energy, and get captured by the condensate star. The accretion process will increase the mass of the condensate star, and may exceed the instability limit, thus causing the collapse of the condensate star. This collapse will probably form black holes from the dark matter star.

In addition to the formation of the dark matter stars during the phase transition epoch in the early universe, dark matter stars may also be formed in another ways. For example, the formation of early stars will make a deep gravitational well to capture the dark matter, which has a high density in the early universe. This process
may result in a condensate dark matter core to form inside normal stars. As more dark matter is captured, due to the self-interaction between particle the accretion rate will be enhanced. The condensate dark matter core will collapse to form black holes once its mass exceeds the instability limit and swallow the
degenerate core of normal matter core, if it exists. This collapse process might have signatures in the observations of high redshift long Gamma-ray Bursts.

As it was pointed out in \cite{ds1,ds2,ds3}, the annihilation of the dark matter particles inside the star can dramatically alter the evolution of the system, by providing a significant source of heating. The dark matter particles annihilation rate is $n_{\chi }^2<\sigma v>$, where the standard annihilation cross section has values of the order of $<\sigma v>\approx 3\times 10^{-26}\;\rm cm^3/s$. The dark matter particle annihilation produces energy at a rate per unit volume of $Q_{DM}=<\sigma v>\rho _{\chi }^2/m_{\chi }$. The annihilation products typically are electrons, photons, and neutrinos. The neutrinos escape the star, while the other annihilation products are trapped in the dark star, and thermalize within the star, eventually heating it up. The luminosity from the dark matter particle annihilation is $L_{DM}=f_Q\int{Q_{DM}dV}$, where $f_Q$ is the fraction of the annihilation energy deposited in the star. Due to their high density, the dark matter condensate stars have a high neutrino luminosity, and this intense neutrino flux may play a significant role in the formation of the gamma ray bursts.

It is also interesting to compare condensate dark matter star to the non-condensate case. In Section IIC, the equation of state for non-condensate dark matter can be described by the equation of state of a collisionless system, $P_{\chi}=\rho_{\chi}c^{2}\sigma_v^{2}$, where $\sigma_v$ is the dimensionless velocity dispersion. This equation of state corresponds to the polytropic model with $n=\infty$. Since the dark matter density in non-condensate state is much smaller than that of the condensate,
the radius of the non-condensate dark matter star must be much larger than that of the condensate star. Therefore we can use the Lane-Emden equation to describe the stellar structure for the non-condensate case, i.e.
\begin{equation}\label{iso}
\frac{1}{r^{2}}\frac{\mathrm{d}}{\mathrm{d}r}\left(\frac{r^{2}}{\rho_{\chi}}\frac{\mathrm{d}\rho_{\chi}}{\mathrm{d}r}\right)=-\frac{4\pi G\rho_{\chi}}{c^{2}\sigma_v^{2}}.
\end{equation}
Eq.~(\ref{iso}), describing the non-condensate case, corresponds to the isothermal sphere, and has an analytical solution given by
\begin{equation}
\rho_{\chi}(r)=\frac{c^{2}\sigma_v^{2}}{2\pi G r^{2}}.
\end{equation}
This solution is unstable and leads to singular behavior of the density near the center. It also does not have a definite boundary with finite mass. This will lead to an extended object made of normal dark matter, while in the condensate case, the dark matter star is very compact.

Alternatively, when normal dark matter is able to form stellar object, it might be described as an adiabatic process, $P_{\chi}\rho_{\chi}^{-\gamma}=\rm const$. In the simplest case, dark matter only has 3 degrees of freedom, i.e. $\gamma=5/3$, and the equation of state is given by
\begin{equation}
P_{\chi}=K\rho_{\chi}^{5/3}.
\end{equation}
The constant $K$ can be determined from the adiabatic invariants,
\begin{equation}
K=P_{\chi}\rho_{\chi}^{-5/3}=P_{c}\rho_{c\chi}^{-5/3}=\frac{\rho_{c\chi}}{m_{\chi}}k_{B}T_{\rm cr}\rho_{c\chi}^{-5/3}=\frac{k_{B}T_{\rm cr}}{m_{\chi}}\rho_{c\chi}^{-2/3},
\end{equation}
where $P_{c}$ and $\rho_{c\chi}$ are central pressure and the density, respectively. The central temperature is taken to be the phase transition temperature. The structure is thus described by the Lane-Emden equation for $n=3/2$. The mass and radius are  given by
\begin{equation}
M=2.714\times4\pi\left(\frac{5K}{8\pi G}\right)^{3/2}\rho_{c\chi}^{1/2},
\end{equation}
and
\begin{equation}
R=3.654\times\left(\frac{5K}{8\pi G}\right)^{1/2}\rho_{c\chi}^{-1/6},
\end{equation}
respectively. For the fiducial values $m_{\chi}=m_{p}$ and $l_{a}=1\;\rm fm$, we obtain $T_{\rm cr}=3.68\times10^8\rm K$. The mass and radius are given by
\begin{eqnarray}
M&=&476.15M_{\odot}\left(\frac{\rho_{c\chi}}{1\rm g/cm^3}\right)^{-1/2},\\
R&=&15.82R_{\odot}\left(\frac{\rho_{c\chi}}{1\rm g/cm^3}\right)^{-1/2}.
\end{eqnarray}
 From these equations it can be clearly seen that for a given mass non-condensate dark matter stars have a much higher radius. This result also justifies the use of the Newtonian Lane-Emden equation to describe this situation.

 If the Bose-Einstein condensation took place in the very early universe, at very high redshifts, as suggested by our analysis of the phase transition, at the moment of star/galaxy formation most, if not all, of the dark matter was already in a condensate form. Hence one may naturally expect the presence of the accreted condensate dark matter inside stellar type objects in the early stages of star and galaxy formation, as well as the formation, through gravitational instabilities, of pure condensate dark matter stars. Thus Bose-Einstein condensate dark matter stars may have been more common in the early Universe than normal dark matter stars.

 \section*{Acknowledgments}

We would like to thank to the anonymous referee for valuable comments and suggestions that helped us to improve our manuscript. This work is
supported by the GRF grants of Hong Kong Government under HKU7010/12P.

\end{document}